\documentclass[noinfoline,stslayout]{imsart} 
\usepackage{amsfonts,amssymb,amsmath,amscd,latexsym}
\usepackage{graphicx}
\usepackage[square,sort,comma,numbers]{natbib}
\usepackage[]{units}
\usepackage{url}
\usepackage{algpseudocode}
\usepackage{qtree}
\usepackage{algorithm}
\newtheorem{exemp}{Example}[section]
\newtheorem{lemma}{Lemma}[section]

\newtheorem{remark}{Remark}

\newcommand\dd{\mathrm{d}}
\newcommand\bigo{\text{O}}

\newcommand\becom{\relax}

\begin{document}

\begin{frontmatter}

\title{Accelerating Metropolis--Hastings algorithms: Delayed acceptance with prefetching}
\runtitle{Delayed acceptance and prefetching for Metropolis--Hastings}

\begin{aug}
 \author{\snm{{\sc Marco Banterle}}}
 \affiliation{Universit{\'e} Paris-Dauphine, CEREMADE, and CREST, Paris}
 \author{\snm{{\sc Clara Grazian}}}
 \affiliation{Sapienza Universit\`a di Roma, Universit{\'e} Paris-Dauphine, CEREMADE, and CREST, Paris}
 \author{\snm{{\sc Christian P.~Robert}}}
 \affiliation{Universit{\'e} Paris-Dauphine, CEREMADE, Dept. of Statistics, University of Warwick, and CREST, Paris}
\end{aug}

%
%
%
%

\begin{abstract} MCMC algorithms such as Metropolis-Hastings algorithms are
slowed down by the computation of complex target distributions as exemplified
by huge datasets.  We offer in this paper an approach to reduce the
computational costs of such algorithms by a simple and universal
divide-and-conquer strategy. The idea behind the generic acceleration is to
divide the acceptance step into several parts, aiming at a major reduction in
computing time that outranks the corresponding reduction in acceptance
probability.  The division decomposes the ``prior x likelihood" term into a
product such that some of its components are much cheaper to compute than
others. Each of the components can be sequentially compared with a uniform
variate, the first rejection signalling that the proposed value is considered no
further, This approach can in turn be accelerated as part of a prefetching
algorithm taking advantage of the parallel abilities of the computer at hand.
We illustrate those accelerating features on a series of toy and realistic
examples.  \end{abstract}

\end{frontmatter}

\section{Introduction}\label{intro}

When running an MCMC sampler such as Metropolis-Hastings algorithms
\citep{robert:casella:2004}, the complexity of the target density required by
the acceptance ratio may lead to severe slow-downs in the execution of the
algorithm. A direct illustration of this difficulty is the simulation from a
posterior distribution involving a large dataset of $n$ points for which the
computing time is at least of order $\bigo(n)$. Several solutions to this issue have
been proposed in the recent literature \citep{korattikara:chen:welling:2013,
neiswanger:wang:xing:2013, scott:etal:2013,wang:dunson:2013}, taking advantage
of the likelihood decomposition 
\begin{equation}\label{eq:baselike}
\prod_{i=1}^n \ell(\theta|x_i) 
\end{equation}
to handle subsets of the data on different processors (CPU), graphical units (GPU), or
even computers.  However, there is no consensus on the method of choice, some
leading to instabilities by removing most prior inputs and others to
approximation delicate to evaluate or even to implement.

Our approach here is to delay acceptance (rather than rejection as in
\cite{tierney:mira:1998}) by sequentially comparing parts of the acceptance
ratio to independent uniforms, in order to stop earlier the computation of the
aforesaid ratio, namely as soon as one term is below the corresponding uniform.
We also propose a further acceleration by combining this idea with
parallelisation through prefetching \citep{brockwell:2006}.

The plan of the paper is as follows: in Section \ref{sec:zero}, we validate the
decomposition of the acceptance step into a sequence of decisions, arguing
about the computational gains brought by this generic modification of
Metropolis-Hastings algorithms and further presenting two toy examples. In
Section \ref{sec:fetch}, we show how the concept of prefetching
\citep{brockwell:2006} can be connected with the above decomposition in order
to gain further efficiency by taking advantage of the parallel capacities of
the computer(s) at hand. Section \ref{sec:examples} study the novel method
within two realistic environments, the first one made of logistic regression
targets using benchmarks found in the earlier prefetching literature and a
second one handling an original analysis of a parametric mixture model via genuine
Jeffreys priors. Section \ref{sec:quatr} concludes the paper.

\section{Breaking acceptance into steps}\label{sec:zero}

In a generic Metropolis-Hastings algorithm the acceptance ratio 
$\nicefrac{\pi(\theta)\,q(\theta,\eta)}{\pi(\eta)q(\eta,\theta)}$ is 
compared with a $\,\mathcal{U}(0,1)$ variate to decide whether or
not the Markov chain switches from the current value $\eta$ to the proposed value $\theta$
\citep{robert:casella:2004}. However, if we decompose the ratio as
an arbitrary product
\begin{equation}\label{eq:prodeck}
{\pi(\theta)\,q(\theta,\eta)}\big/{\pi(\eta)q(\eta,\theta)} = \prod_{k=1}^d \rho_k(\eta,\theta)\,,
\end{equation}
where the only constraint is that the functions $\rho_k$ are all positive
and accept the move with probability
\begin{equation}\label{eq:prodike}
\prod_{k=1}^d \min\left\{\rho_k(\eta,\theta),1\right\}\,,
\end{equation}
i.e.~by successively comparing uniform variates $u_k$ to the terms $\rho_k(\eta,\theta)$,
the same target density $\pi(\theta)$ is stationary for the resulting Markov chain. 
In practice, sequentially comparing those probabilities with uniform variates means
that the comparisons stop at the first rejection, meaning a gain in computing time
if the most costly items are kept till the final comparisons.

\begin{lemma}
The Metropolis-Hastings algorithm with acceptance probability \eqref{eq:prodike}
has the same stationary distribution as the original Metropolis-Hastings 
algorithm with acceptance probability \eqref{eq:prodeck}.
\end{lemma}

The mathematical validation of this simple if surprising result can be seen as
a consequence of \cite{christen:fox:2005}. This paper reexamines
\cite{fox:nicholls:1997}, where the idea of testing for acceptance using an
approximation and before computing the exact likelihood was first suggested.
In \cite{christen:fox:2005}, the original proposal density $q$ is used to
generate a value $\theta'$ that is tested against an approximate target
$\pi^0$. If accepted, $\theta'$ is then tested against the true target $\pi$,
using a pseudo-proposal $q^0$ that is simply reproducing the earlier
preliminary step. The validation in \cite{christen:fox:2005} follows from
standard detailed balance arguments. Indeed, take an arbitrary decomposition of
the joint density on $(\theta,\eta)$ into a product,
$$
\pi(\theta)\,q(\theta,\eta) = \omega_1(\theta) \,q(\theta,\eta) 
\prod_{k=2}^d \omega_k(\theta)\,,
$$
associated with \eqref{eq:prodike}. Then
\begin{align*}
\pi(\eta) q(\eta,\theta) &\prod_{k=1}^d \min\{\rho_k(\eta,\theta),1\} \\
& =\min\{\omega_1(\theta) \,q(\theta,\eta),\omega_1(\eta) \,q(\eta,\theta)\} 
\prod_{k=1}^d \min\{\omega_k(\theta),\omega_k(\eta)\}\\
&= \pi(\theta) q(\theta,\eta) \prod_{k=1}^d \min\{\rho_k(\theta,\eta),1\}
\end{align*}
which is symmetric in $(\theta,\eta)$, hence establishes 
the detailed balance condition \citep{tierney:1994,robert:casella:2004}.

\becom
\begin{remark}
Note that the validation can also be derived as follows: accepting or rejecting
a proposal using the first ratio is equivalent to generating from the ``prior",
then having used this part of the target as the new proposal leads to its
cancellation from the next acceptance probability and so on. 
\end{remark}

\begin{remark}
While the purpose of \cite{doucet:pitt:deligiannidis:kohn:2014} is
fundamentally orthogonal to ours, a special case of this decomposition of the
acceptance step in the Metropolis--Hastings algorithm can be found therein.
In order to achieve a manageable
bound on the convergence of a particle MCMC algorithm, the authors decompose
the acceptance in a Metropolis--Hastings part based on the parameter of
interest and a second Metropolis--Hastings part based on an auxiliary variable.
They then demonstrate stationarity for the target distribution in this modified
Metropolis--Hastings algorithm.
\end{remark}

\begin{remark}
Another point of relevance is that this modification of the acceptance probability
in the Metropolis--Hastings algorithm cannot be expressed as an unbiased estimator
of the likelihood function, which would make it a special case of pseudo-marginal
algorithm \citep{andrieu:roberts:2009}.
\end{remark}

The delayed acceptance scheme found in \cite{fox:nicholls:1997} efficiently
reduces the computing cost only when the approximation is good enough since the
probability of acceptance of a proposed value is smaller in this case.  In
other words, the original Metropolis-Hastings kernel dominates the new one in
Peskun's \citep{peskun:1973} sense. The most relevant question raised by
\cite{christen:fox:2005} is how to achieve a proper approximation, but in our
perspective a natural approximation is obtained by breaking original data in
subsamples and considering the corresponding likelihood part.


\begin{figure}
\begin{center}
\includegraphics[width=.4\textwidth]{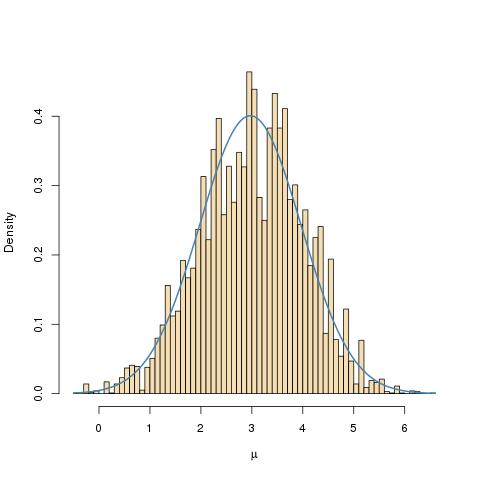}
\end{center}
\caption{\label{fig:norlayed}
Fit of a two-step Metropolis--Hastings algorithm applied to a normal-normal
posterior distribution $\mu|x\sim N(x/(\{1+\sigma_\mu^{-2}\},1/\{1+\sigma_\mu^{-2}\})$
when $x=3$ and $\sigma_\mu=10$, based on $T=10^4$ iterations and a first acceptance
step considering the likelihood ratio and a second acceptance step considering the prior ratio,
resulting in an overall acceptance rate of 12\%
}\end{figure}


\begin{figure}
\includegraphics[width=.3\textwidth]{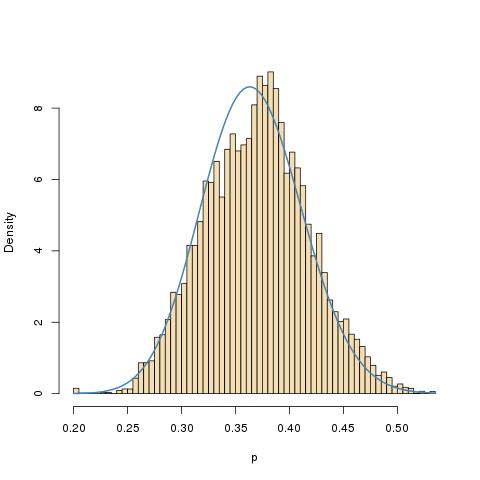}
\includegraphics[width=.3\textwidth]{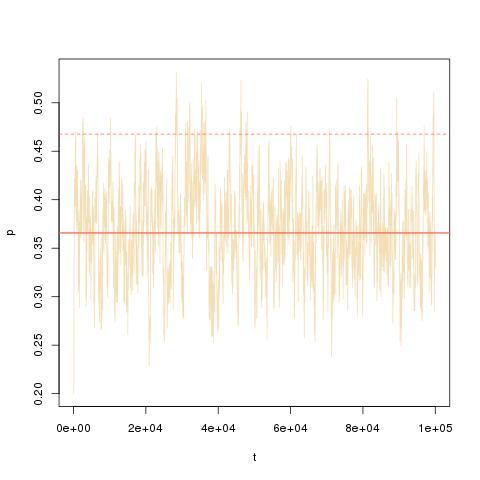}
\includegraphics[width=.3\textwidth]{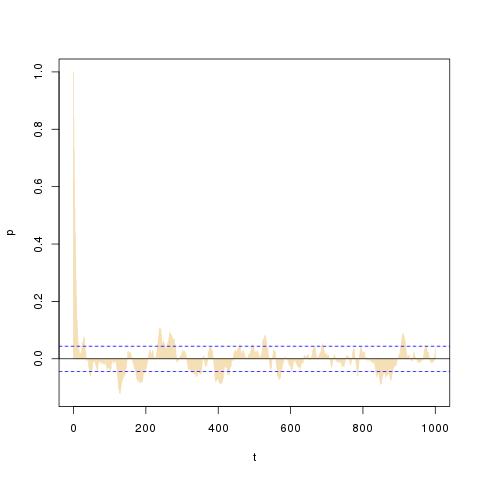}
\caption{\label{fig:binomial}
(left)
Fit of a multiple-step Metropolis--Hastings algorithm applied to a Beta-binomial
posterior distribution $p|x\sim Be(x+a,n+b-x)$ when $N=100$, $x=32$, $a=7.5$    
and $b=.5$. The binomial $\mathcal{B}(N,p)$ likelihood is replaced with a product of $100$ Bernoulli terms and
an acceptance step is considered for the ratio of each term. The histogram is based on $10^5$
iterations, with an overall acceptance rate of 9\%; 
(centre) raw sequence of successive values of $p$ in the Markov chain simulated 
in the above experiment; (right) autocorrelogram of the above sequence.
}\end{figure}

A primary application of this result which pertains to all Bayesian applications
is to separate the likelihood ratio from the prior ratio and to compare only
one term at a time with a corresponding uniform variate. For instance, when
using a costly prior distribution (as illustrated in Section \ref{sec:mixis} in the
case of mixtures), the first acceptance step is solely based on the ratio of
the likelihoods, while the second acceptance probability involves the ratio of
the priors, which does not require to be computed when the first step leads to
rejection. Most often, though, the converse decomposition applies to complex or
just costly likelihood functions, in that the prior ratio may first be used to
eliminate values of the parameter that are too unlikely for the prior density.
As shown in Figure 
\ref{fig:norlayed}, a standard normal-normal example confirms that the true posterior and the
histogram resulting from such a simulated sample are in agreement.

In more complex settings, the above principle also applies to a product of
individual terms as in a likelihood so each individual likelihood can be
evaluated separately. This approach increases both the variability of the
evaluation and the potential for rejection, but, if each likelihood term is
sufficiently costly to compute the decomposition brings some improvement in
execution time. The graphs in Figure \ref{fig:binomial} illustrate
an implementation of this perspective in the Beta-binomial case, namely when
the binomial $\mathcal{B}(N,p)$ observation $x=32$ is replaced with a sequence of $N$ Bernoulli
observations. The fit is adequate on $10^5$ iterations, but the autocorrelation
in the sequence is very high (note that the ACF is for the 100 times thinned
sequence) while the acceptance rate falls down to 9\%. (When the original
$y=32$ observation is (artificially) divided into 10, 20, 50, and 100 parts,
the acceptance rates are 0.29, 0.25, 0.12, and 0.09, respectively.)

\becom\begin{remark}
We stress that the result remains valid even when the likelihood function or the prior are not integrable
over the parameter space. Therefore the prior may well be improper. For instance, when the prior distribution
is constant, a two-stage acceptance scheme reverts to the original one.
\end{remark}

\begin{remark}\label{rem:order}
Another point worth noting is that the order in which the product \eqref{eq:prodike}
is explored is irrelevant, since all terms need be evaluated for a Nev value to be
accepted. It therefore makes sense to try to optimise this order by
considering ranking the entries according to the success rate so far, starting with the
least successful values. An alternative is to rank according to the last computed values of
the likelihood at each datapoint, as (a) those values are available for the last accepted proposal
and (b) it is more efficient to start with the highest likelihood values. Even though this form
of reordering seems to contradict the fundamental requirement for Markovianity and hence ergodicity of the resulting
MCMC algorithm, reordering has no impact on the overall convergence of the resulting
Markov chain, since an acceptance of a proposal does require computing all likelihood values,
while it does or should improve the execution speed of the algorithm. Note however that specific
decompositions of the product may lead to very low acceptance rates, for instance when picking only
outliers in a given group of observations.
\end{remark}

While the delayed acceptance methodology is intended to cater to complex likelihoods or priors, it does not bring
a solution {\em per se} to the ``Big Data'' problem in that (a) all terms in the product must eventually be computed;
(b) the previous terms (i.e., those computed for the last accepted value of the parameter) 
must all be stored in preparation for comparison or recomputed; 
(c) the method is ultimately inefficient for very large datasets, unless blocks of observations are
considered together. The following section addresses more directly the issue of large datasets.

\becom
As a final remark, we stress the analogy between our delayed acceptance algorithm
and slice sampling \citep{neal:1997,robert:casella:2004}.  Based on the same decomposition 
\eqref{eq:baselike}, slice sampling proceeds as follows
\begin{enumerate}
\item simulate $u_1,\ldots,u_n\sim\mathcal{U}(0,1)$ 
and set $\lambda_i=u_i\ell(\theta|x_i)$ $(i=1,\ldots,n)$;
\item simulating $\theta^\prime$ as a uniform under the constraints
$\ell_i(\theta^\prime|x_i)\ge \lambda_i$ $(i=1,\ldots,n)$.
\end{enumerate}
to compare with delayed sampling which conversely
\begin{enumerate}
\item simulate $\theta^\prime\sim q(\theta^\prime|\theta)$;
\item simulate $u_1,\ldots,u_n\sim\mathcal{U}(0,1)$
and set $\lambda_i=u_i\ell(\theta|x_i)$ $(i=1,\ldots,n)$;
\item check that
$\ell_i(\theta^\prime|x_i)\ge \lambda_i$ $(i=1,\ldots,n)$.
\end{enumerate}
The differences between both schemes are thus that (a) slice sampling always accepts a move,
(b) slice sampling requires the simulation of $\theta^\prime$ under the constraints, which may
prove unfeasible, and (c) delayed
sampling re-simulates the uniform variates in the event of a rejection. In this respect, delayed sampling appears
as a ``poor man's" slice sampler in that values of $\theta's$ are proposed until one is accepted.

\section{Parallelisation and prefetching}\label{sec:fetch}

\subsection{The concept of prefetching}
\label{sec:consix}

\begin{figure}
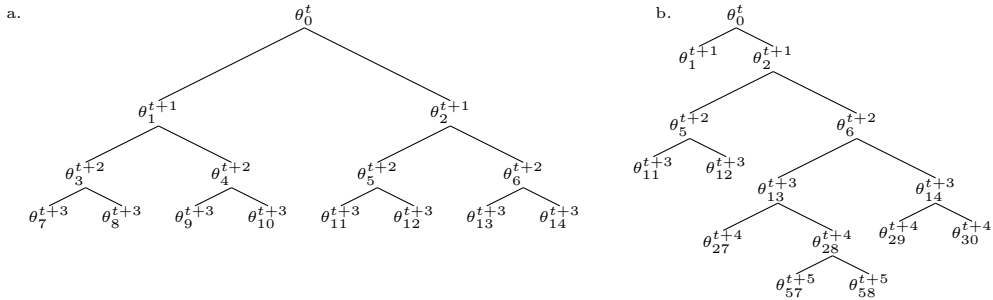


\qtreecenterfalse

\tiny{
a.
\Tree [.$\theta_0^t$ 
		[.$\theta_1^{t+1}$ 
			[.$\theta_3^{t+2}$ 
				$\theta_{7}^{t+3}$  $\theta_{8}^{t+3}$
			]
			[.$\theta_4^{t+2}$ 
				$\theta_{9}^{t+3}$ $\theta_{10}^{t+3}$			
			]		
		] 
		[.$\theta_2^{t+1}$ 
			[.$\theta_5^{t+2}$ 
				$\theta_{11}^{t+3}$  $\theta_{12}^{t+3}$
			]
			[.$\theta_6^{t+2}$ 
				$\theta_{13}^{t+3}$ $\theta_{14}^{t+3}$			
			]
	    ]
	  ]
	  \hskip 1cm  	   
b.
\Tree [.$\theta_0^t$ 
		[.$\theta_1^{t+1}$ ] 
		[.$\theta_2^{t+1}$ 
			[.$\theta_5^{t+2}$ 
				$\theta_{11}^{t+3}$ 
				$\theta_{12}^{t+3}$	
							] 
			[.$\theta_6^{t+2}$ 
				[.$\theta_{13}^{t+3}$ 
					$\theta_{27}^{t+4}$ 
					[.$\theta_{28}^{t+4}$
						$\theta_{57}^{t+5}$ 
						$\theta_{58}^{t+5}$
					]
				]
				[.$\theta_{14}^{t+3}$ 
					$\theta_{29}^{t+4}$ 
					$\theta_{30}^{t+4}$
				] 	!\qsetw{0.8cm}		
			]   !\qsetw{2.5cm}
		] !\qsetw{1cm}
	  ]

}	    
\normalsize
\caption{\label{fig:tree} 
(a) Example of a tree generated by static prefetching \citep{brockwell:2006} requiring 6 additional processors;
(b) example of a dynamic-prefetched tree \citep{strid:2010} for the same number of processors.}
\end{figure}

\normalsize
Prefetching, as defined by \cite{brockwell:2006}, is a programming method that
accelerates the convergence of a single MCMC chain by making use of parallel
processing to compute posterior values ahead of time.  Consider a
generic random-walk Metropolis-Hastings algorithm and say the chain reached
time $t$; as shown in Figure \ref{fig:tree}.a the subsequent steps can be
represented by a binary decision tree, where at time $t+k$ the chain has $2^k$
\emph{possible} future states, with $2^{k-1}$ different posterior values
(rejection events, represented by convention by odd subscripts, share the posterior value 
taken by their parent).

In a parallel environment, given that the value of the target at
$\theta_0^t$ is already computed and that the master thread evaluates the
target for $\theta_2^{t+1}$, the most na{\"\i}ve version of prefetching
\citep{brockwell:2006} requires $\sum_{i=1}^{k} 2^{i-1}$ additional
threads to compute all possible target values for the subsequent $k$ steps.
After collecting all results, the master thread proceeds to accept/reject
over those $k$ steps in a standard sequential manner at the cost of just one step
if we consider the evaluation of the target as being the most expensive part.

\becom
The static prefetching algorithm proceed as in Algorithm \ref{algo:pref_base}, in the setting
of Figure \ref{fig:tree} (a) and the call to $K=7$ cores.

\becom\begin{algorithm}
\caption{Prefetching algorithm}
\label{algo:pref_base}

\emph{When the chain is in state $\theta_0^t$ at time $t$:}\hfill\break
\begin{enumerate}
\item (serial) Construct the tour $$ T= \{ \theta_{i_1}, \dots \, , \theta_{i_K} \} = \{ \theta_2^{t+1},\theta_4^{t+2},\theta_6^{t+2},\theta_8^{t+3},\theta_{10}^{t+3},\theta_{12}^{t+3},\theta_{14}^{t+3} \}$$
of all \emph{possible} future values of $\theta$ for the next 3 steps;
\item Scatter $\theta_{i_k}$, $k=1, \dots\,, K$ among all available cores;
\item (parallel) Core $k$ compute $\pi(\theta_{i_k})$;
\item Collect all the computed $\pi(\theta_{i_k})$;
\item (serial) Run the Metropolis--Hasting scheme as usual until the end of the tour $T$;
\item Update $t$ accordingly and set $\theta_0^t$ as the last reached value.
\end{enumerate}
\end{algorithm}

A fundamental requirement for validating this scheme is that both the sequence
of random variables $(\zeta_t)$ underlying the random-walk steps and the
sequence of uniform $u_t \sim U(0,1)$ driving the accept/reject steps need be
simulated beforehand and remain identical across all leaves at an iteration
$t$. Thus, all proposals at time $t+k$ will be generated (regardless of the
starting point) using a single $\zeta_{t+k}$ and possibly tested for acceptance
using a single $u_{t+k}$.

\begin{remark}
This requirement is trivially implemented for random-walk Metropolis--Hastings
algorithms, where the proposal satisfies $\theta_{t+1} = \theta_t +
h(\zeta_{t+1})$. It actually applies for all Metropolis--Hastings algorithms,
from independent proposals to more elaborate (Riemannian) Hamiltonian Monte
Carlo \citep{neal:2012,girolami:2011}.
\end{remark}

How far ahead in time we can push this scheme is clearly limited by the number
of available (additional) processors. It is worth stressing that a single
branch of the computed tree is eventually selected by the method, resulting in
a quite substantial waste of computing power, even though Rao--Blackwellisation
\cite{jacob:robert:smith:2010} could recycle the other branches towards
improved estimation.

More efficient approaches towards the exploration of the above tree have been
proposed for example by \cite{strid:2010} and \cite{angelino:etal:2014} by
better guessing the probabilities of exploration of both the acceptance and the
rejection branches at each node.  Define $\gamma_{2i_k+2} = \gamma_{i_k} \times
\alpha_{i_k}$ the probability of visiting $\theta_{2i_k+2}$ given the
probability of reaching its parent ($\gamma_{i_k}$) and an estimation of the
probability of accepting it ($\alpha_{i_k}$) (the case of rejection is easily
derived).  The basic static prefetching follows from defining $\alpha_{i_k} =
0.5$ ; better schemes can thus be easily devised, taking advantage of (i) the
observed acceptance rate of the chain ($\alpha_{i_k} = \alpha_{obs}$), (ii) the
sequence of already stored uniform variates ($\alpha_{i_k}$ being the average
probability of acceptance given $u_t$), and (iii) any available (fast)
approximation of the posterior distribution ($\alpha_{i_k} = \text{Pr}\left(
u_t < \nicefrac{\hat{\pi}(\theta_{i_k})}{\hat{\pi}(\theta_{i_{k-1}})}
\right)$), towards pursuing or abandoning the exploration of a given branch and
thus increasing the expected number of draws per iteration (see Figure
\ref{fig:tree}(b)).

\becom
Algorithm \ref{algo:tour_const} formalises this advance by detailing point 1. of Algorithm \ref{algo:pref_base},
 where $\gamma_{i_k}$ depends on the probability of reaching the parent node
$\gamma_{\lfloor (i_k-1)/2\rfloor}$ and on its probability to be accepted
$\alpha_{\lfloor (i_k-1)/2\rfloor}$.  $\alpha_{i_k}$ is then the only thing
determining the type of prefetching used, the most basic static prefetching
being when $\alpha_{i_k} = 0.5$ 
 
\begin{algorithm}
\caption{Tour Construction}
\label{algo:tour_const}

\begin{enumerate}
\item Set $T = \{ \theta_{i_1} \} = \{ \theta_2^{t+1} \}$, add $\theta_1$ to the candidates and assign it probability $\gamma_1$\\
	
\hspace{-1cm} \emph{For $k = 2, \dots \, , K$} do:
\begin{enumerate}
		\item Add to the candidate points the children of $i_{k-1} \rightarrow \{ 2i_{k-1}+1 , 2i_{k-1}+2 \}$;
		\item Assign them probability $\gamma_{ 2i_{k-1}+1 }$ and $\gamma_{ 2i_{k-1}+2 }$;
		\item Select the candidate with the highest probability and add it to the tour.
\end{enumerate}
\end{enumerate}
\end{algorithm}

As an illustration of the above, assume the chain has reached the state $\theta_0^t$ and that two processors are available.
The first one is forced to compute the target value at $\theta_2^{t+1}$ as it obviously stands next in line.
The second one can then be employed to compute the value of the target density at $\theta_4^{t+2}$
(anticipating a rejection) or at $\theta_6^{t+2}$ (anticipating instead an acceptance). A few remarks are in order:
\begin{enumerate}
\renewcommand{\theenumi}{(\alph{enumi})}
        \item in the basic static case, both possibilities are symmetric;
        \item if one takes into account the observed acceptance rate (equal, say, the golden standard
        \citep{gelman:gilks:roberts:1996} of $\alpha = 0.234$) then preparing for a rejection is more appealing;
        \item however, we can also exploit the prior knowledge of the next uniform, say $u_{t} \approx 10^{-4}$, 
	hence computing instead $\theta_6^{t+2}$ is a safer strategy;
        \item at last, if an approximation of the target distribution $\hat\pi(\cdot)$ is almost freely available, 
	we may compute $\hat{\pi}(\theta_0^t)/\hat{\pi}(\theta_2^{t+1})$ and base our decision on that approximate
	ratio.
\end{enumerate}
Assuming more processors are at our disposal, the method can iterate, starting over
from the last chosen point and its children, but still considering all the values
examined up till then as candidates. Take for instance a setting of
8 processors. We now follow strategy (b) above and keep the the notation 
($\theta_{i_k}, \gamma_{i_k}$) for representing (\emph{proposed point}, \emph{probability of exploration}).

The tour (made of the points selected to prefetch) starts with
$$
\mathcal{T} = \{ (\theta_{2}^{t+1}, 1) , \; (\theta_4^{t+2}, (1-\alpha)=0.766) \}
$$ 
and the corresponding set of candidate points is $\mathcal{C} = \{ (\theta_{6}^{t+2}, 0.234) \}$.
At the following move, in order to allocate the next processor, we have to add to the candidates 
the points resulting from both an acceptance and a rejection of $\theta_4^{t+2}$, so:
$$ 
\mathcal{C} = \{ (\theta_{6}^{t+2}, 0.234), \; (\theta_{8}^{t+3}, 
0.766 \times 0.766 = 0.588), \;(\theta_{10}^{t+3}, 0.766 \times 0.234 = 0.179 ) \} 
$$ 
The candidate $(\theta_{8}^{t+3}, 0.766 \times 0.766 = 0.588)$ is the one with maximum probability.
It is thus removed from $\mathcal{C}$ and inserted into $\mathcal{T}$ as the next point to be computed.
Similarly, the following three steps are such that rejection of moves from the last chosen point 
is the most probable setting. When 6 cores are exploited, the corresponding two sets are given by

\begin{align*}
\mathcal{T} =& \{ (\theta_{2}^{t+1}, 1) , \; (\theta_4^{t+2},0.77), \, (\theta_{8}^{t+3},0.59), 
\; (\theta_{16}^{t+4}, 0.45), \; (\theta_{32}^{t+5}, 0.34) , \; (\theta_{64}^{t+6}, 0.26) \} \\
\mathcal{C} =& \{ (\theta_{6}^{t+2}, 0.23), \; (\theta_{10}^{t+3}, 0.18 ), \; (\theta_{18}^{t+4}, 0.14), 
\; (\theta_{34}^{t+5}, 0.10) , \; (\theta_{66}^{t+6}, 0.081) \}
\end{align*}

\noindent The next two candidates points stemming from $(\theta_{64}^{t+6},
0.264) $ are $(\theta_{130}^{t+7}, 0.062)$ and $(\theta_{128}^{t+7}, 0.202)$.
Thus, branching the tree from the top and computing $(\theta_{6}^{t+2}, 0.234)$
is more likely to yield an extra useful (i.e. 	involved in future ratio computations) point. Hence, $(\theta_{6}^{t+2}, 0.234)$ is 
added to the tour $\mathcal{T}$.  At the final step, $(\theta_{128}^{t+7},
0.202)$ is eventually selected for this move and the tree is at last completed for all 8 cores,
returning

\begin{align*}
\mathcal{T} = \{ &(\theta_{2}^{t+1},1) , \; (\theta_4^{t+2},0.77), \, (\theta_{8}^{t+3},0.59), \; (\theta_{16}^{t+4},0.45),\\
 &(\theta_{32}^{t+5},0.34) , \; (\theta_{64}^{t+6},0.26) , \; (\theta_{6}^{t+2},0.23), \; (\theta_{128}^{t+7},0.20) \}.
\end{align*}
 
\noindent Interested readers are referred to \cite{strid:2010} for a detailed illustration of other prefetching strategies.

\subsection{Prefetching and Delayed Acceptance}
\label{sec:prefDA}

Combining the technique of delayed acceptance as described in Section \ref{sec:zero} with the above
methodology of prefetching is both natural and straightforward, with major prospects in terms of
improved performances.

Assume for simplicity's sake that the acceptance ratio breaks as follows:
\begin{equation}\label{eq:MH_rate_2split}
{\pi(\theta)\,q(\theta,\eta)}\big/{\pi(\eta)q(\eta,\theta)} =
\rho_1(\eta,\theta) \times \rho_2(\eta,\theta) \end{equation} where the
evaluation of $\rho_1(\eta,\theta)$ is inexpensive relative to the one of
$\rho_2(\eta,\theta)$.  We can then delay computing in parallel the values of
$\rho_2(\eta,\theta)$ and use instead $\rho_1(\eta,\theta)$ to help the
prefetching algorithm in constructing the tour.  By early rejecting a proposed
value at step $k$ due to the event $u_1^{t+k} > \rho_1(\eta,\theta)$, we can
immediately cut the corresponding branch, thus reaching further in depth into
the tree without the need for extra processors.  An algorithmic representation
of this fusion is provided in Algorithm \ref{algo:pref+DA},
where $u_{(i_k)}$ is the reference uniform given $t$ and the depth in the tree of
$\theta_{i_k}$.  Note that all the elements resulting from a rejection have the
same target value as their parent, which thus need not be recomputed.


\begin{algorithm}
\caption{Tour Construction with Delayed Acceptance}
\label{algo:pref+DA}

\begin{enumerate}
\item Set $k=1$, $T = \{ \theta_{i_k} \} = \{ \theta_2^{t+1} \}$, 
add $\theta_1$ to the candidates and assign it probability $\gamma_1$\\
	\begin{enumerate}
	\item[(a)]  \textbf{while}( $u_{(i_k)} > \rho_1(\theta_{(i_k-\cdot)/2}, \theta_{i_k})$ )
		 set $i_k = 2i_k+1$ (rejection), add $\theta_{i_k}$ to the tour;
	\end{enumerate}
\noindent \emph{For $k = 2, \dots \, , K$} do:
	\begin{enumerate}
		\item[(b)] Add to the candidate points the children of $i_{k-1} \rightarrow \{ 2i_{k-1}+1 , 2i_{k-1}+2 \}$;
		\item[(c)] Assign them probability $\gamma_{ 2i_{k-1}+1 }$ and $\gamma_{ 2i_{k-1}+2 }$;
		\item[(d)] Select the candidates with highest probability ($\theta_{i_{max}}$), \\ add it to the tour and set $i_k = i_{max}$;
		\item[(e)] \textbf{while}( $u_{(i_k)} > \rho_1(\theta_{(i_k-\cdot)/2}, \theta_{i_k})$ )
		 set $i_k = 2i_k+1$ (rejection), add $\theta_{i_k}$ to the tour.
\end{enumerate}
\end{enumerate}
\end{algorithm}

Moreover if our target density is a posterior distribution, written as
$$
\pi(\theta) \propto p(\theta) \times \prod_{i=1}^n \ell(\theta|x_i) \,,
$$
where $p(\cdot)$ is a computationally cheap prior and $\ell(\cdot)$ is an
expensive individual likelihood, we can split the acceptance ratio in
\eqref{eq:MH_rate_2split} as 
\begin{equation}\label{eq:MH_rate_lik_split}
\dfrac{\pi(\theta)\,q(\theta,\eta)}{\pi(\eta)q(\eta,\theta)} =
\dfrac{p(\theta) \, q(\theta,\eta) \, \prod_{i=1}^r \ell(\theta|x_i) }{ p(\eta)
\, q(\eta,\theta) \, \prod_{i=1}^r \ell(\eta|x_i) } \times
\dfrac{\prod_{i=r+1}^n \ell(\theta|x_i)}{\prod_{i=r+1}^n \ell(\eta|x_i)}
\end{equation}
where $1 < r \ll n$. Then, making use of the fact that $\prod_{i=1}^r \ell(\theta|x_i)$ produces a 
``free lunch" (if biased) estimator of $\prod_{i=r+1}^n
\ell(\theta|x_i)$ and hence a subsequent estimator of $\rho_2(\theta,\eta)$, we can directly set 
 $\alpha_\theta = \mathbb{I}_{u_2^{t+k} \leq \hat{\rho}_2(\eta,\theta)}$ 
in the tour construction in order to find our most likely path to follow in the decision tree, 
or else exploit this decomposition by setting the probability $\alpha_\theta = \min \{ \beta,\hat{\rho}_2(\eta,\theta) \}$ 
with $\beta \in (0,1]$.

When the approximation $\hat{\rho}_2(\eta,\theta)$ is good enough, both these
strategies have been shown to yield the largest efficiency gains
in \cite{strid:2010}.  \cite{angelino:etal:2014} propose another more involved
prefetching procedure that uses an approximation of the target with better
performances.

\begin{remark}
Generalising this basic setting to different product decompositions of the
acceptance rate and/or to settings with more terms in the product is
straightforward and above remarks about delayed acceptance, in particular
Remark \ref{rem:order}, still hold.
\end{remark}

\begin{remark} Since, for stability reasons, the log-likelihood is often the
computed quantity, this may prohibit an easy derivation of an {\em unbiased
estimator} of $\rho_2(\cdot)$ by sub-sampling techniques. Nonetheless, an
estimated $\hat{\rho}_2(\cdot)$ can be used by the prefetching algorithm to
construct a speculative tour as it does not contribute to any expression
involving the actual chain. As mentioned above, a poor approximation could clearly lower the
performance improvement but there is no consequence on the actual convergence
of the chain. \end{remark}

All the examples were coded in C++.  The logistic examples are run on a cluster
composed of 12-cores (Intel\textsuperscript{\textregistered}
Xeon\textsuperscript{\textregistered} CPU @ 2.40GHz) nodes, using up to 4 nodes
for a total of 48 cores, and make use of
Open-MPI\footnote{\url{http://www.open-mpi.org/}} for communications between
cores.  The mixture example is run on a 8 cores
(Intel\textsuperscript{\textregistered} Xeon\textsuperscript{\textregistered}
CPU @ 3.20GHz) single machine, using OpenMP\footnote{\url{http://openmp.org/}}
for parallelisation.

\section{Examples}\label{sec:examples}

To illustrate the improvement brought by our conjunction of delayed acceptance
and prefetching, we study two different realistic settings to reflect on the generality
of the method. First, we consider a Bayesian analysis of a logistic regression
model, on both simulated and real datasets, to assess the computational gain
brought by our approach in a ``BigData" environment where obtaining the
likelihood is the main computational burden. Then we investigate a mixture
model where a formal Jeffreys prior is used, as it is not available in
closed-form and does require an expensive approximation by numerical or Monte
Carlo means. This constitutes a realistic example of a setting where the prior
distribution is a burdensome object, even for small dataset.

\subsection{Logistic Regression}

While a simple model, or maybe exactly because of that, logistic regression is
widely used in applied statistics, especially in classification problems.  The
challenge in the Bayesian analysis of this model is not generic, since simple
Markov Chain Monte Carlo techniques providing satisfactory approximations, but
stems from the data-size itself. This explains why this model is used as a
benchmark in some of the recent accelerating papers
\citep{korattikara:chen:welling:2013, neiswanger:wang:xing:2013,
scott:etal:2013,wang:dunson:2013}. Indeed, in ``big Data" setups, MCMC is
deemed to be progressively inefficient and researchers are striving for
progresses in keeping it effective even in this scenario, focusing mainly on
parallel computing and on sub-sampling but also on tweaking the classic
Metropolis scheme itself. Our proposal contributes to those attempts.

\subsubsection{Synthetic Data}
\label{sec:log_synth}
Simulated data gives us an opportunity to play with complexity and computing costs in
a perfectly controlled manner. In the setting of a logistic regression model, we have
for instance incorporated a controlled amount of ``white" computation
proportional to the number of observations to mimic the case of a truly expansive
likelihood function. In practice, those datasets are made of $n=1000$ observations
with 5 covariates.

In this setup, we tested various variants of dynamic prefetching and came to
the conclusion that, beyond a generic similarity between the approaches, the
best results relate to a probability $\min \{ \beta,\hat{\rho}_2(\eta,\theta)
\}$ of picking a branch at each node, especially as this allows (contrarily to
picking the most likely path) for branched trees. Our strategy is as detailed
in Section \ref{sec:prefDA}; we split the data (and the likelihood) into two
sets, one being considerately smaller than the other, and we implement 
delayed acceptance, splitting between the diffuse Gaussian prior, the elliptical normal
proposal, and the smallest fraction of the likelihood on the one side, and the expensive
part of the likelihood on the other side.

\begin{figure}
\begin{center}
\includegraphics[width=.8\textwidth]{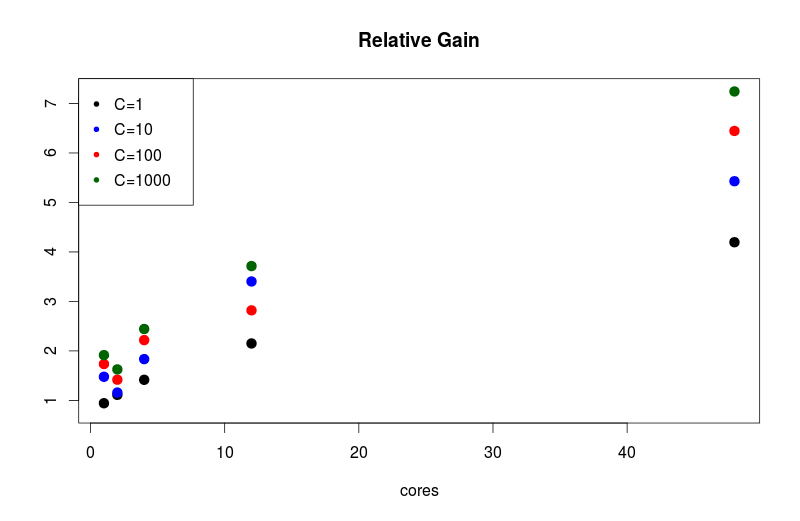}
\end{center}
\caption{\label{fig:log_synth}
Relative gain ($RG$) of the delayed acceptance method over a classic Metropolis--Hastings, both in combination with prefetching. Each colour represent a different order of artificial cost for the likelihood.
}\end{figure}

We compare in Figure \ref{fig:log_synth} a condensed index of runtime and output quality defined as 
$$ 
RG = { \left( \frac{ESS_{DA}}{t_{DA}}\right) }\Big/{ \left( \frac{ESS_{MH}}{t_{MH}}\right) } 
$$ 
where $t$ is the average runtime of the algorithm, $ESS$ is the Effective
Sample Size and $DA$ and $MH$ stand for delayed acceptance and
standard Metropolis--Hastings, respectively. The graph covers combinations of basic MCMC, prefetched MCMC
with different numbers of cores and delayed acceptance, each version with a
range of different orders $C$ of (artificial) computational cost for the baseline likelihood
function.  Each chain is the product of $T=10^5$ iterations with a burn-in of
$10^3$ iterations. The proposal distribution is a Gaussian random walk with covariance matrix 
equal to the asymptotic covariance of the MLE estimator.

\begin{table}
\begin{center}
    \begin{tabular}{ | l | c | c | c | c | c |}
    \hline
    \textbf{Algorithm} & \textbf{ESS} (aver.) & \textbf{ESS} (sd) & \textbf{$\tau$} (aver.) & \textbf{Acceptance rate} (aver.) & \textbf{Acceptance rate} (sd) \\ \hline
     MH & 18595.33 & 2011.067 & 5.3777 & 0.2577 & 0.0295 \\ \hline
	 MH with DA & 14062.19 & 4430.08 & 7.1112 & 0.2062 & 0.05405 \\ \hline    
    \end{tabular}
\end{center}
    \caption{Comparison of effective sample sizes (ESS), autocorrelation time ($\tau$) and acceptance rates for 5 repetitions of the simulated logistic experiment, averaged over number of cores used and likelihood cost $C$ (both of which should not influence the output quality of the algorithm but just the computing time), for delayed acceptance and standard Metropolis--Hastings} \label{tab:log_synth}
\end{table}

The results in Figure \ref{fig:log_synth} and Table \ref{tab:log_synth}
highlight how our combination of prefetching and delayed acceptance improves
upon the basic algorithm. Indeed, even though the acceptance rate of the
delayed acceptance chain is lower than the regular chain, Table \ref{tab:log_synth}, as soon as the likelihood computation starts to be costly the algorithm
brings a gain up to two times the number of independent samples per unit of
time.  The logarithmic behaviour of the gain with respect to the number of
cores used is well known in the prefetching literature \citep{strid:2010}, but
we note how for increasing costs of the likelihood delayed acceptance
becomes more and more efficient.

\begin{figure}
\centerline{\includegraphics[scale=0.3]{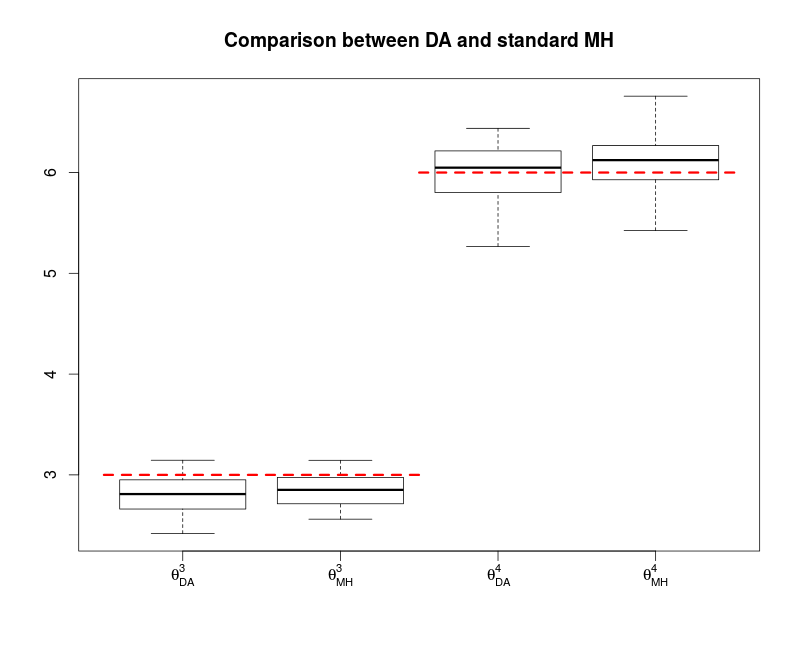}}
\caption{Boxplot representing the approximations to the posterior distribution of two parameters 
of a logistic regression model with 5 covariates (simulated data).}
\label{boxplot_logistic}
\end{figure}

Figure \ref{boxplot_logistic} details the MCMC outcome in the first case. The
box-plots represent approximations to the marginal posterior distributions of
two coefficients in a logistic regression based on simulated data. Those
box-plots are quite similar in terms of both mean values and variability when
comparing the standard Metropolis--Hastings algorithm with the standard
Metropolis--Hastings algorithm with delayed acceptance. Both approximations are
concentrated around the true values of the parameters.

\subsubsection{Higgs Boson Data}
\label{sec:log_higgs}

ATLAS (A Toroidal LHC Apparatus) is a particle detector experiment constructed
at the Large Hadron Collider (LHC), at CERN, and designed to search for new
particles based on collisions of protons of extraordinarily high energy. The
ATLAS experiment has recently observed \citep{ATLAS:2013} evidence for the
Higgs boson decaying into two tau particles, but the background noise is
considerably high.  The ATLAS team then launched a machine learning challenge
through the Kaggle\footnote{ \url{https://www.kaggle.com} } depository,
providing public data in the form of simulations reproducing the behaviour of
the ATLAS experiment. This training dataset is made of 25,0000 events with 30
feature columns and a binary label stating if the event is a real signal of
decay or just background noise. Modelling this dataset through a logit model is
thus adequate.

We compared the combined algorithm of delayed acceptance and prefetching from
Section \ref{sec:log_synth} with regular dynamic prefetching as in \cite{strid:2010},
both with $T=10^6$ iterations after $10^4$ burn-in iterations. The portion of
the sample used in the first acceptance ratio (and used to approximate the
remaining likelihood at the next step) is 5\%, i.e. 12500 points. The proposal
for the logit coefficients is again a normal distribution with covariance
matrix the asymptotic covariance of the MLE estimator, obtained from a subset
of the data, and adapted during the burn-in phase.

In this experiment, we used $48$ cores in parallel and obtained that the
delayed acceptance algorithm runs almost 6 times faster than its classic
counterpart, in line with the average number of draws per iteration obtained
(159.22 for delayed acceptance versus just 29.03 for the standard version).
The acceptance rate, although quite low for both the algorithms, remained
constant through repetitions for both the methods, namely around 0.5\%,
and so was the relative ESS, around $10^{-4}$. 

Once again, the approximations obtained with both algorithms are very close. In
this case, a huge sample size leads to highly concentrated posterior
distributions.  We have chosen to represent only the posteriors of two
parameters, for clarity's sake, although all other parameters show a substantial
similarity between both algorithms.

\begin{figure}
\centerline{\includegraphics[scale=0.3]{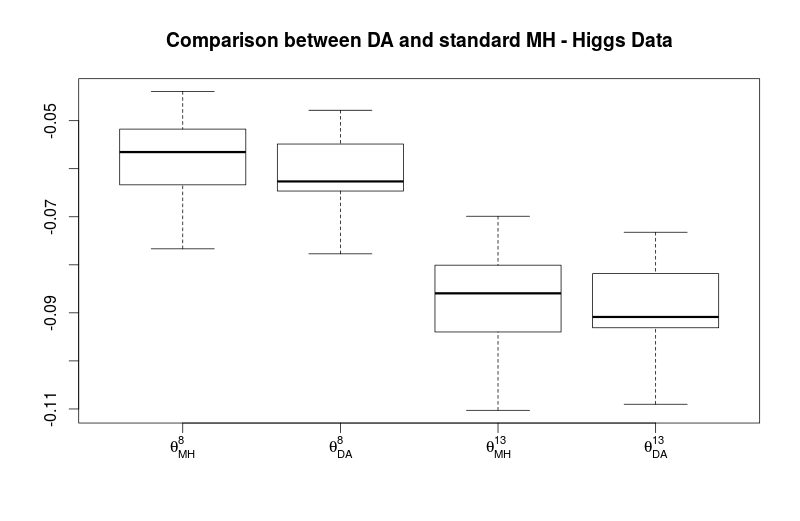}}
\caption{Boxplot representing the approximations to the posterior distribution of two parameters of the logistic regression model with 30 covariates used with the Higgs Boson Data.}
\label{boxplot_higgs}
\end{figure}

\subsection{Mixture Model}\label{sec:mixis}

Consider a standard mixture model \citep{maclachlan:peel:2000} with a fixed number of components
\begin{equation}\label{eq:theMix}
\sum_{i=1}^k w_i\,f(x|\theta_i)\,,\quad \sum_{i=1}^k w_i=1\,.
\end{equation}
This standard setting nonetheless offers a computational challenge in that the reference objective Bayesian approach 
based on the Fisher information and the associated Jeffreys prior \citep{jeffreys:1939,robert:2001}
is not readily available for computational reasons and has thus not being implemented so far. Proxys
using the Jeffreys priors of the component of \eqref{eq:theMix} have been proposed instead in the past, with the
drawback that since they always lead to improper posteriors, ad hoc corrections had to implemented
\citep{diebolt:robert:1994,roeder:wasserman:1997,stephens:1997}.

When relying instead on dependent improper priors, it is less clear that the
improperness of the posterior distribution happens. For instance, we consider
the genuine Jeffreys prior for the complete set of parameters in
\eqref{eq:theMix}, derived from the Fisher information matrix for the whole
model. While establishing the analytical properness of the associated posterior
is beyond the goal of the current paper {\em (work in progress)}, we handle
large enough samples to posit that a sufficient number of observations is
allocated to each component and hence the likelihood function dominates the
prior distribution. (In the event the posterior remains improper, the
associated MCMC algorithm should exhibit a transient behaviour.)

We therefore argue this is an appropriate and realistic example for implementing
delayed acceptance since the computation of the prior density is clearly costly, 
relying on many integrals of the form:
\begin{equation} 
-\int_{\mathcal{X}} \frac{\partial^2 \log \left[\sum_{i=1}^k
w_i\,f(x|\theta_i)\right]}{\partial \theta_h \partial
\theta_j}\left[\sum_{i=1}^k w_i\,f(x|\theta_i)\right] \dd x \,.
\end{equation}
Indeed, integrals of this form cannot be computed analytically and thus their
derivation involve numerical or Monte Carlo integration. We are therefore in a
setting where the prior ratio---as opposed to the more common case of the
likelihood ratio---is the costly part of the target evaluated in the
Metropolis-Hastings acceptance ratio. Moreover, since the Jeffreys prior
involves a determinant, there is no easy way to split the computation further
than "prior times likelihood". Indeed, the likelihood function is
straightforward to compute. Hence, the delayed acceptance algorithm can be
applied by simply splitting between the prior $p^J(\psi)$ and the likelihood
$\ell(\psi|x)$ ratios, the later being computed first.
Since in this setting the prior is (according to simulation studies) improper,
picking the acceptance ratio at the second step solely on the prior
distribution may create trapping states in practice, even though the method
would remain valid.  We therefore opt to the stabilising alternative to keep a
small fraction (chosen to be 2\% in our implementation) of the likelihood to
regularise this second acceptance ratio by multiplication with the prior.  
This choice translates into Algorithm \ref{algo:mix-algo}.

\begin{algorithm}
\caption{Metropolis-Hastings with Delayed Acceptance for Mixture Models}
\label{algo:mix-algo}
Set $\ell_2(\cdot|x) = \sum \limits_{i=1}^{ \left\lfloor 0.02 n \right\rfloor } \ell(\cdot|x_i)$ and  $\ell_1(\cdot|x) = \sum \limits_{i=\left\lfloor 0.02 n \right\rfloor+1}^{n} \ell(\cdot|x_i)$\hfill\break
\begin{enumerate}
\item Simulate $\psi^\prime\sim q(\psi^\prime|\psi)$;
\item Simulate $u_1,u_2\sim\mathcal{U}(0,1)$ and set $\lambda_1=u_1\ell_1(\psi|x)$;
\item \textbf{if} $\ell_1(\psi^\prime|x)\le \lambda_1$, repeat the current parameter value and return to 1; \\
\textbf{else} set $\lambda_2=u_2 \ell_2(\psi|x) p^J(\psi)$; 
\item \textbf{if} $\ell_2(\psi^\prime|x) p^J(\psi^\prime)\ge \lambda_2$ accept $\psi^\prime$; \\ \textbf{else} repeat the current parameter value and return to 1.
\end{enumerate}
\end{algorithm}

\begin{figure}
\includegraphics[scale=0.5]{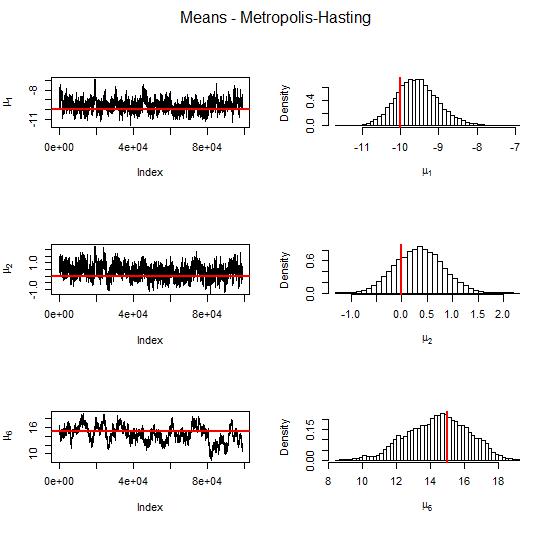}
\caption{Marginal posteriors on the means of a 3-components Gaussian mixture. 
The sample is made of 1000 observations, with true values indicated by red vertical
lines. The MCMC output is obtained via Metropolis-Hastings, for $10^5$ simulations.}
\label{postmeansMH}
\end{figure}

\begin{figure}
\includegraphics[scale=0.5]{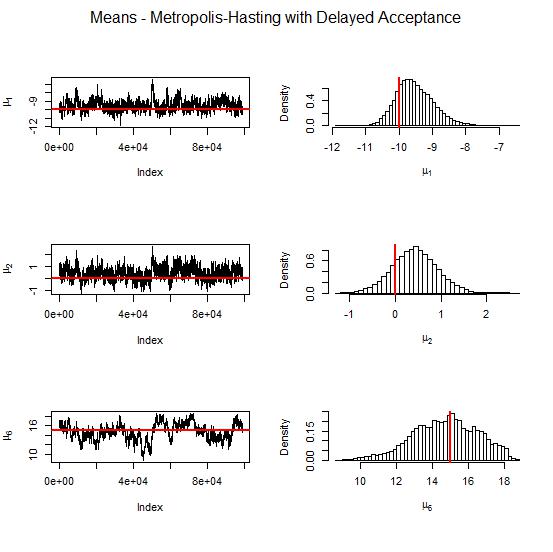}
\caption{Same graph as above when supplemented by a delayed acceptance step.}
\label{postmeansDA}
\end{figure}

\begin{figure}
\includegraphics[scale=0.5]{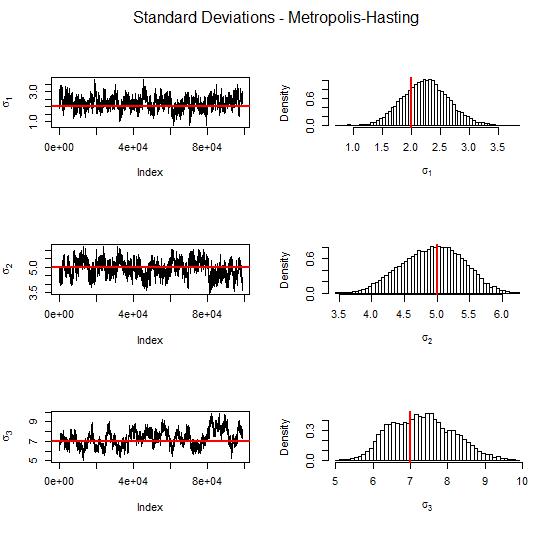}
\caption{Marginal posteriors on the standard deviations of a 3-components Gaussian mixture.
The sample is made of 1000 observations, with true values indicated by red vertical
lines. The MCMC output is obtained via Metropolis-Hastings, for $10^5$ simulations.}
\label{postsdMH}
\end{figure}

\begin{figure}
\includegraphics[scale=0.5]{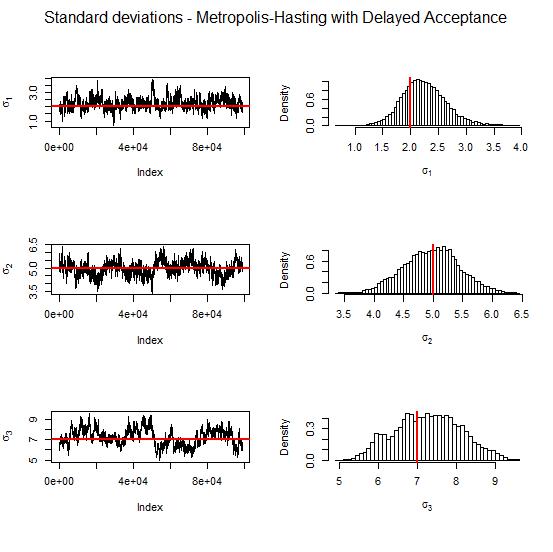}
\caption{Same graph as above when supplemented by a delayed acceptance step.}
\label{postsdDA}
\end{figure}

\begin{figure}
\includegraphics[scale=0.5]{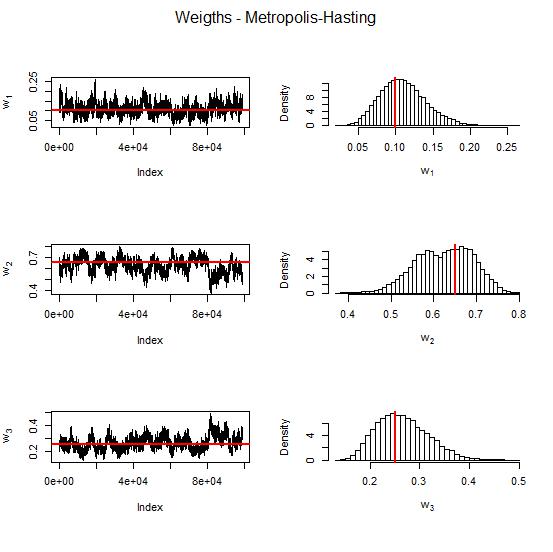}
\caption{Marginal posteriors on the weights of a 3-components Gaussian mixture.
The sample is made of 1000 observations, with true values indicated by red vertical
lines. The MCMC output is obtained via Metropolis-Hastings, for $10^5$ simulations.}
\label{postwwMH}
\end{figure}

\begin{figure}
\includegraphics[scale=0.5]{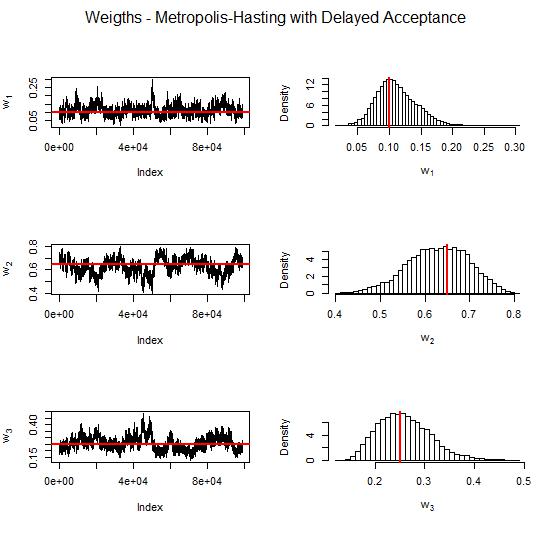}
\caption{Same graph as above when supplemented by a delayed acceptance step.}
\label{postwwDA}
\end{figure}

An experiment comparing a standard Metropolis--Hastings implementation with a
Metropolis--Hastings version relying on delayed acceptance (again, with and without
prefetching) is summarised in Table \ref{tab:Mix} and in Figures
\ref{postmeansMH}--\ref{postwwDA}.  When implementing the prefetching option we
have only resorted to a maximum of 8 processors (for availability reasons).  
Data was simulated from the following Gaussian mixture model: 
\begin{equation}
f(y|\theta)=0.10\mathcal{N}(-10,2)+0.65\mathcal{N}(0,5)+0.25\mathcal{N}(15,5). 
\end{equation}
\label{eq:mixmodel}

The graphs in Figures \ref{postmeansMH}--\ref{postwwDA} report on the resulting approximations to the marginal posterior
distributions of the parameters of a three-component Gaussian mixture, obtained
with a Metropolis-Hastings algorithm in the first instances and with the
delayed acceptance algorithm in the second instance (in both cases, in
conjunction with prefetching). Both raw MCMC sequences and histograms show a
remarkable similarity in their approximation to the posterior distributions. In
particular, the estimation of the parameters of the third component (which has
the highest variance) shows the same highest variability in both the cases. As
an aside we also notice that label switching does occur there
\citep{jasra:holmes:stephens:2005}.

\begin{table}
\begin{center}
    \begin{tabular}{ | l | c | c | c | c | c |}
    \hline
    \textbf{Algorithm} & \textbf{ESS} (aver.) & \textbf{ESS} (var) & \textbf{time} (aver.) & \textbf{time} (var) & \textbf{Acceptance rate}\\ \hline
     MH & 168.85 & 62.37 & 517.60 & 0.51 & 0.50 \\ \hline
	 MH + DA & 112.74 & 2155.89 & 322.38 & 28.69 & 0.43\\ \hline    
	 MH + pref. & 173.30 & 506.57 & 225.18 & 0.03 & 0.50 \\ \hline    
	 MH + DA + pref. & 150.18 & 841.06 & 192.65 & 76.28 & 0.43 \\ \hline    
    \end{tabular}
\end{center}
    \caption{Comparison between different performance indicators 
for four algorithmic implementations in the example of mixture estimation, based
on 5 replicas of the experiments according to model \eqref{eq:mixmodel} with a sample size $n=1,000$, 
$10^5$ simulations and a burnin of $10^4$  simulations.}
\label{tab:Mix}
\end{table}

\becom

The delayed acceptance algorithm naturally exhibit a smaller acceptance rate
(around $45\%)$ when compared with a standard MCMC algorithm with no
parallelisation (around $50\%$), but the difference is minor. 

Furthermore, this lesser efficiency is to be balanced by the major improvement
that the computational time is $1.6$ times less with MCMC with delayed
acceptance with respect to a standard MCMC with to  for $10^5$ simulations and
a sample size $n=1000$, while the gain is even higher while using prefetching
(the computational time is reduced two times).  The average number of draws
obtained when using delayed acceptance with prefetching is $7.52$ with respect
to $2.9$ when using only prefetching, giving the same indication that the
reduction in computing time. 

Overall, the accepted values are highly correlated in all the cases; in
particular, when using the delayed acceptance algorithm the effective sample
size is around $1.5$ times less than when using a standard
Metropolis--Hastings. The computational time of the delayed acceptance
algorithm with prefetching is 3 times less than the standard MCMC, against a
less strong reduction of the effective sample.

\section{Conclusion}\label{sec:quatr}
While the choice of splitting the target distribution into pieces ultimately depends on the respective
costs of computing the said pieces and of reducing the overall acceptance rate, this generic alternative to 
the standard Metropolis--Hastings approach should be considered on a customary basis since it requires
very little modification in programming and since it can be tested against the basic version.

The delayed acceptance algorithm presented in \eqref{sec:zero} could broadly
decrease the computational time \textit{per se}; a counterweight is the reduced
acceptance rate, nevertheless the examples presented in Section
\ref{sec:examples} suggest that the gain in terms of computational time is not
linear with respect to the reduction of the acceptance rate. 

Furthermore, our delayed acceptance algorithm does naturally merge with the
widening range of prefetching techniques, in order to make use of
parallelisation and reduce the overall computational time even more
significantly. 

Most settings of interest are open to take advantage of the proposed method, if
mostly when either the likelihood or the prior distribution are costly to
evaluate. The setting when the likelihood function can be factorised in an useful
way represents the best gain brought by our solution, in terms of computational time, as it mainly
exploits the parallelisation. 

\subsubsection*{Acknowledgments}

Thanks to Christophe Andrieu for very helpful advice in validating the
multi-step acceptance procedure. The massive help provided by Jean-Michel Marin
and Pierre Pudlo towards an implementation on a large cluster has been 
fundamental in the completion of this work. Christian P. Robert research is partly
financed by Agence Nationale de la Recherche (ANR, 212, rue de Bercy 75012
Paris) on the 2012--2015 ANR-11-BS01-0010 grant ``Calibration'' and by a
2010--2015 senior chair grant of Institut Universitaire de France.



\end{document}